\documentclass[journal]{IEEEtran}

\usepackage[T1]{fontenc}
\usepackage{cite}
\usepackage{amsmath,amssymb,bm}
\usepackage{graphicx}
\usepackage{xcolor}
\usepackage{siunitx}
\usepackage{algorithm}
\usepackage{algpseudocode}
\usepackage{booktabs}
\usepackage{hyperref}

\hypersetup{
  colorlinks=true,
  linkcolor=blue,
  citecolor=blue,
  urlcolor=blue
}

\begin{document}

\title{Adaptive Eigenvector Continuation for Full-Vector Photonic Waveguide Mode Emulation}

\author{Daniel~Rodr\'iguez-Guill\'en and~Lorena~Vel\'azquez-Ibarra
	
	\thanks{This work was supported in part by Secretaría de Ciencia, Humanidades, Tecnología e Innovación (SECIHTI, Ref. CBF-2025-I-2037). D.R.-G. acknowledges financial support from SECIHTI (Graduate Fellowship, CVU-1142156). (\textit{Corresponding author: L. Vel\'azquez-Ibarra.})}

	\thanks{D. Rodr\'iguez-Guill\'en and L. Vel\'azquez-Ibarra are with the Departmento de F\'isica, Divisi\'on de Ciencias e Ingenier\'ias, Universidad de Guanajuato, Le\'on 37150, M\'exico (e-mail: d.rodriguezguillen@ugto.mx; \mbox{lorenav@fisica.ugto.mx).}}
}

\markboth{Journal of \LaTeX\ Class Files,~Vol.~XX, No.~XX, Month~2026}%
{Rodr\'iguez-Guill\'en: Adaptive Eigenvector Continuation for Fast Full-Vector Photonic Waveguide Design}

\maketitle

\begin{abstract}
Photonic waveguide design often requires repeated full-vector Maxwell eigenmode solves over wavelength, geometry, and material parameters. We present an adaptive eigenvector-continuation framework for accelerating and stabilizing these modal sweeps. The method constructs a reduced basis from selected full-order modal snapshots, solves projected Maxwell eigenproblems at new query points, reconstructs the modal fields, and monitors accuracy with a full operator residual. We demonstrate three regimes. In fixed-geometry wavelength sweeps of a strip waveguide, well-distributed snapshots reproduce the target modal branch with low residual and low effective-index error. In a multimode ridge waveguide, a shared reduced basis containing several modal families enables robust broadband mode-family tracking and residual-guided adaptive enrichment. In geometry-dependent width sweeps, the method gives accurate effective-index predictions and high field overlap, but the residual reveals moving-boundary errors caused by non-smooth changes of the discrete operator on a fixed Cartesian grid. These results show that adaptive eigenvector continuation is an operator-consistent modal emulator and diagnostic tool for photonic waveguide sweeps.

\end{abstract}


\section{Introduction}
Photonic waveguide design relies on repeated numerical solutions of Maxwell eigenmode problems. Integrated waveguides, photonic crystal fibers, and nonlinear or quantum photonic devices are commonly optimized over wavelength, geometry, and material parameters in order to control quantities such as the effective index, propagation constant, group index, chromatic dispersion, effective mode area, nonlinear coefficient, birefringence, and phase-matching conditions. These quantities are central to dispersion engineering, nonlinear frequency conversion, supercontinuum generation, and photon-pair source design~\cite{Turner06,Dudley06,FrancisJones2018,Velazquez19}. However, obtaining them with high fidelity usually requires solving a large sparse full-vector eigenvalue problem at every point of the parameter sweep.

This repeated solve workflow can become computationally expensive when high spatial resolution, vectorial field accuracy, dispersive materials, and accurate dielectric-interface treatment are required. Moreover, broadband or multimode sweeps introduce an additional difficulty: the eigenmodes computed independently at different wavelengths or design points must be consistently assigned to the same physical mode family. In practice, repeated eigensolves can reorder modes, miss higher-order branches, or lose modal continuity when effective indices become close, when polarization mixing occurs, or when the field profiles evolve significantly across the sweep.

A common way to reduce the computational cost is to solve the full-order problem only at a limited number of points and then interpolate the effective index or propagation constant. Although this approach is simple and inexpensive, it does not enforce Maxwell's equations at the query point. This limitation becomes especially important when derivatives of the propagation constant are required, because small interpolation errors in the propagation constant can be amplified in group index, group velocity dispersion, and phase matching calculations. Therefore, nonlinear and quantum photonic design workflows benefit from reduced models that are not only fast, but also consistent with the underlying Maxwell operator.

Reduced order models~\cite{Quarteroni2015,Rozza2008} address this need by constructing low-dimensional approximations from selected high-fidelity solutions. In this context, eigenvector continuation (EC) provides a projection-based strategy for parameter-dependent eigenvalue problems~\cite{Frame2018ECSubspace,DuguetColloquiumEC}. Instead of interpolating scalar modal quantities directly, EC builds a reduced basis from full-order eigenvector snapshots computed at selected training points. At a new query point, the Maxwell eigenproblem is projected onto this reduced subspace, and a small generalized eigenvalue problem is solved. The resulting approximation is therefore obtained from the discretized Maxwell operator evaluated at the query point, providing an operator-consistent reduced modal emulator.

In this work, we develop an adaptive eigenvector-continuation framework for full-vector photonic waveguide eigenmode problems. The proposed workflow constructs reduced bases from selected full-order modal snapshots, solves projected Maxwell eigenproblems for new wavelength or design points, reconstructs the approximate modal fields, and monitors the quality of the reduced solution using a full-operator residual. The method is formulated independently of the particular full-vector field representation, allowing the reduced basis to be built from different Maxwell eigenmode formulations as long as the corresponding discretized operator can be projected consistently.

We demonstrate the method in three representative scenarios. First, we consider fixed-geometry wavelength sweeps of a single-mode strip waveguide, showing how the EC basis reproduces the fundamental modal branch and how snapshot placement affects the residual. Second, we apply the method to a multimode ridge waveguide, where several modal branches are embedded in the reduced basis. This case demonstrates the usefulness of EC for robust mode-family tracking, since the reduced space contains the targeted modal families and reduces the fragility of independent eigensolves. Third, we study geometry-dependent sweeps, where the waveguide width is varied. In this case, EC provides accurate effective-index estimates and high field overlap, but the residual can remain large because moving dielectric boundaries on a fixed Cartesian grid induce non-smooth changes in the discrete operator.

\section{Full-Vectorial Maxwell Eigenmode Formulation}

We consider source-free, time-harmonic Maxwell's equations with an $\exp(-i\omega t)$ convention,
\begin{align}
\nabla\times \mathbf{E} &= i\omega \mu(\lambda)\mathbf{H}, \\
\nabla\times \mathbf{H} &= -i\omega \epsilon(\lambda)\mathbf{E},
\end{align}
where $\epsilon(\lambda)$ denotes a wavelength-dependent permittivity (e.g., a Sellmeier model) and $\mu$ is typically taken as constant in non-magnetic media. For waveguides invariant in the longitudinal direction $z$, guided modes are assumed to propagate as $\exp(i\beta z)$, where $\beta$ is the propagation constant. Eliminating either $\mathbf{E}$ or $\mathbf{H}$ yields a curl-curl-type eigenmode formulation for the transverse field components, with $\beta$ (or $\beta^2$ depending on the chosen formulation) as the eigenvalue.

After discretizing the transverse cross-section using any consistent full-vectorial discretization scheme~\cite{Fallahkhair2008}, the eigenmode computation becomes a large sparse eigenvalue problem. In general, Maxwell discretizations naturally yield a generalized eigenvalue problem \cite{Kappesser2023RBMMaxwell},
\begin{equation}
A(\lambda;\theta)\,\vec{x}(\lambda;\theta) \;=\; \beta(\lambda;\theta)\, B(\lambda;\theta)\,\vec{x}(\lambda;\theta),
\label{eq:gevp_beta}
\end{equation}
where $\vec{x}$ is the discrete eigenvector containing the transverse field degrees of freedom, and $A$ and $B$ are sparse matrices produced by the discretization. The parameter vector $\theta$ collects the geometrical degrees of freedom as well as material parameters defining the permittivity distribution. The explicit dependence on $\lambda$ is essential for dispersive materials because $\epsilon(\lambda)$ and $k_0(\lambda)=2\pi/\lambda$ modify the coefficients of the operator across the sweep.

A broadband characterization requires evaluating \(\beta(\lambda)\) across a wavelength grid. In this work, the full-order sweep is used as the reference solution, while EC is used to approximate selected
modal families from a reduced snapshot basis. For notational simplicity, we write the generalized eigenvalue problem using \(\beta\) as the eigenvalue. The same EC construction applies to formulations where the discrete eigenvalue is \(\beta^2\), provided that the projected problem and the residual are defined consistently with the chosen full-order formulation.

\section{Adaptive Eigenvector Continuation for Maxwell Eigenmodes}

Eigenvector continuation is a projection-based reduced-subspace method for parameter-dependent eigenvalue problems~\cite{Frame2018ECSubspace,DuguetColloquiumEC}. While EC is frequently introduced in Hamiltonian settings, its mathematical mechanism is more general: it relies on the smooth (often analytic) dependence of the operators on parameters, which implies that the corresponding eigenvectors vary smoothly with parameters except near isolated singularities associated with eigenvalue collisions in the complexified parameter space \cite{SarkarLeeConvergenceEC}. This viewpoint explains why a small set of eigenvector ``snapshots'' computed at selected training points can span a low-dimensional subspace that accurately represents solutions over a wide parameter range, including extrapolative regimes \cite{Frame2018ECSubspace}.

In the Maxwell waveguide eigenproblem~\eqref{eq:gevp_beta}, the parameter dependence arises from wavelength $\lambda$ (through $k_0$ and $\epsilon(\lambda)$) and from geometry/material parameters $\theta$. We compute high-fidelity eigenvectors at a small set of training points $\{(\lambda_i,\theta_i)\}_{i=1}^{n_b}$, using mode tracking only on this reduced set to ensure that all snapshots correspond to the same targeted mode family. Let $\{\vec{x}_i\}_{i=1}^{n_b}$ denote the resulting snapshots and define the snapshot matrix
\begin{equation}
V = [\vec{x}_1,\vec{x}_2,\ldots,\vec{x}_{n_b}].
\end{equation}
Here, \(N\) denotes the number of full-order degrees of freedom of the discretized Maxwell eigenproblem, so that \(\vec{x}_i \in \mathbb{C}^{N}\). The integer \(n_b\) is the number of basis vectors retained in the EC subspace. For a single-mode basis, \(n_b\) is equal to the number of snapshot parameter points. For a multimode basis, several eigenvectors
can be added at each snapshot point. For a new query point $(\lambda,\theta)$, EC approximates the eigenvector as
\begin{equation}
\vec{x}(\lambda,\theta) \approx V\,\vec{y}(\lambda,\theta),
\end{equation}
where $\vec{y}\in\mathbb{C}^{n_b}$ are reduced coefficients. Substituting into~\eqref{eq:gevp_beta} and applying a Galerkin projection onto the EC subspace yields the reduced generalized eigenvalue problem
\begin{equation}
A_r(\lambda,\theta)\,\vec{y}(\lambda,\theta) \;=\; \beta_r(\lambda,\theta)\, B_r(\lambda,\theta)\,\vec{y}(\lambda,\theta),
\label{eq:reduced_gevp}
\end{equation}
with reduced matrices
\begin{align}
A_r(\lambda,\theta) &= V^{H} A(\lambda,\theta) V, \label{eq:projection1}\\
B_r(\lambda,\theta) &= V^{H} B(\lambda,\theta) V.
\label{eq:projection2}
\end{align}
Since \(n_b \ll N\), the projected eigenproblem in \eqref{eq:reduced_gevp} is much smaller than the original sparse eigenproblem. However, in the present implementation, the online stage still requires evaluating the full-order Maxwell operator at the query point and projecting it onto the EC basis through \eqref{eq:projection1} and \eqref{eq:projection2}. Thus, the reduced eigenvalue solve itself is inexpensive, but operator assembly and projection can remain the dominant online cost.

The approximate full field is reconstructed as
\(\vec{x}_r = V\vec{y}\), and the approximate propagation constant is \(\beta_r\). Unlike direct interpolation of \(\beta(\lambda)\), EC enforces the Maxwell operator structure
at each query point through the projected eigenproblem. This operator-consistent construction is especially useful when derived quantities depend on \(\beta(\lambda)\), its derivatives, or on the reconstructed modal fields.

To make the procedure adaptive, we monitor an error indicator at each query point. We choose a residual norm associated with the full operator evaluated on the reduced solution,
\begin{equation}
r(\lambda,\theta) = \frac{\left\|A(\lambda,\theta)\vec{x}_r - \beta_r B(\lambda,\theta)\vec{x}_r\right\|}{\left\|\beta_r B(\lambda,\theta)\vec{x}_r\right\|},
\end{equation}
and we enrich the basis by adding new snapshot points when $r(\lambda,\theta)$ exceeds a prescribed tolerance. 
The adaptive procedure used in the numerical examples is summarized in Table~\ref{tab:adaptive_ec}. In all examples, the snapshot matrix is orthonormalized before projection to improve the conditioning of the reduced matrices.

\begin{table}[t]
\centering
\caption{Adaptive EC sweep.}
\label{tab:adaptive_ec}
\begin{tabular}{p{0.08\linewidth}p{0.84\linewidth}}
\hline
Step & Operation \\
\hline
1 & Choose an initial snapshot set
\(\mathcal{P}_0=\{(\lambda_i,\theta_i)\}\). \\
2 & Solve the full-order eigenproblem at each snapshot point and
assemble \(V=[\vec{x}_1,\ldots,\vec{x}_{n_b}]\). \\
3 & Orthonormalize or compress \(V\) to improve conditioning. \\
4 & For each query point, form
\(A_r=V^H A(\lambda,\theta)V\) and
\(B_r=V^H B(\lambda,\theta)V\). \\
5 & Solve \(A_r\vec{y}=\beta_r B_r\vec{y}\) and reconstruct
\(\vec{x}_r=V\vec{y}\). \\
6 & Compute the full-operator residual \(r(\lambda,\theta)\). \\
7 & If \(\max r\) is below tolerance, accept the sweep. Otherwise,
add a full-order snapshot at the largest-residual query point and
repeat. \\
\hline
\end{tabular}
\end{table}

\section{Fixed-Geometry Wavelength Sweeps}

We first evaluate the proposed EC framework in fixed-geometry wavelength sweeps. In this setting, the transverse waveguide cross-section is unchanged and the parameter dependence enters through the wavelength-dependent Maxwell operator, including the material dispersion. This provides a clean regime for EC because the dielectric interface remains fixed on the computational grid and the targeted modal family varies smoothly with wavelength.

This section addresses two questions. First, how does the placement of the snapshots affect the quality of the reduced basis? Second, how well do the residual and effective-index error diagnose the accuracy of the EC reconstruction? These questions are studied using a single-mode strip waveguide, which provides a simple benchmark before considering multimode tracking and geometry-dependent sweeps.

\subsection{Single-mode strip waveguide benchmark}

\begin{figure}[h]
    \centering
    \includegraphics[width=0.8\linewidth]{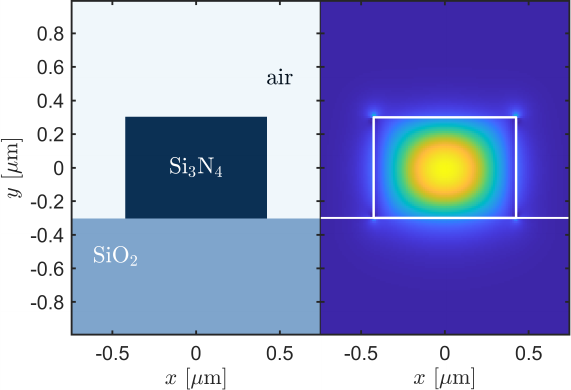}
    \caption{Single-mode strip-waveguide design. Left: Si\(_3\)N\(_4\) strip waveguide on a SiO\(_2\) substrate with air cladding. Right: field-intensity profile of the fundamental quasi-TE mode used as the target modal family for the EC basis.}
    \label{Fig1:Design}
\end{figure}

As a first benchmark, we consider a Si\(_3\)N\(_4\) strip waveguide on a SiO\(_2\) substrate with air cladding. The core width and height are \(0.85~\mu\mathrm{m}\) and \(0.60~\mu\mathrm{m}\), respectively. The computational window is \(1.5 \times 2.0~\mu\mathrm{m}^2\), discretized with \(N_x=133\) and \(N_y=178\) grid points. Dirichlet boundary conditions are imposed at the transverse boundaries, and tensorial subpixel smoothing is used to represent the dielectric interface~\cite{Farjadpour2006,Kottke2008}. Figure~\ref{Fig1:Design} shows the waveguide cross-section and the fundamental quasi-TE mode used as the target mode.

A full-order wavelength sweep is performed over the interval
\[
\lambda \in [0.8,1.6]~\mu\mathrm{m},
\]
and is used as the reference solution. The EC bases are then constructed from selected full-order snapshots of the same mode. The purpose of this benchmark is not only to reproduce the effective-index curve, but also to quantify how the location and number of snapshots affect the full-operator residual and the effective-index error.

\subsection{Snapshot-placement strategies}

The accuracy of EC depends on how well the snapshot basis spans the modal subspace over the target parameter interval. To study this effect, we compare several snapshot-placement strategies. These strategies are chosen to distinguish between endpoint sampling, local sampling around the center of the interval, clustered sampling near one side of the interval, and approximately uniform sampling across the full wavelength range.

The tested snapshot sets are summarized in Table~\ref{tab:strip_snapshot_sets}. Each set defines the columns of the EC basis,
\[
V = [x(\lambda_1),x(\lambda_2),\ldots,x(\lambda_{n_b})],
\]
where \(x(\lambda_i)\) is the full-order eigenvector of the fundamental quasi-TE mode at the snapshot wavelength \(\lambda_i\).

\begin{table}[t]
\centering
\caption{Snapshot-placement strategies used in the single-mode strip-waveguide benchmark.}
\label{tab:strip_snapshot_sets}
\begin{tabular}{ll}
\hline
Strategy & Snapshot wavelengths \((\mu\mathrm{m})\) \\
\hline
Endpoints & \(0.8,\;1.6\) \\
Endpoints + center & \(0.8,\;1.2,\;1.6\) \\
Three local points & \(1.1,\;1.2,\;1.3\) \\
Five local points & \(1.0,\;1.1,\;1.2,\;1.3,\;1.4\) \\
Left cluster & \(0.8,\;0.9,\;1.0,\;1.1\) \\
Right cluster & \(1.3,\;1.4,\;1.5,\;1.6\) \\
Five uniform points & \(0.8,\;1.0,\;1.2,\;1.4,\;1.6\) \\
Seven uniform points & \(0.8,\;0.9,\;1.05,\;1.2,\;1.35,\;1.5,\;1.6\) \\
\hline
\end{tabular}
\end{table}

\subsection{Effective-index reconstruction}

\begin{figure}[h]
    \centering
    \includegraphics[width=1.03\linewidth]{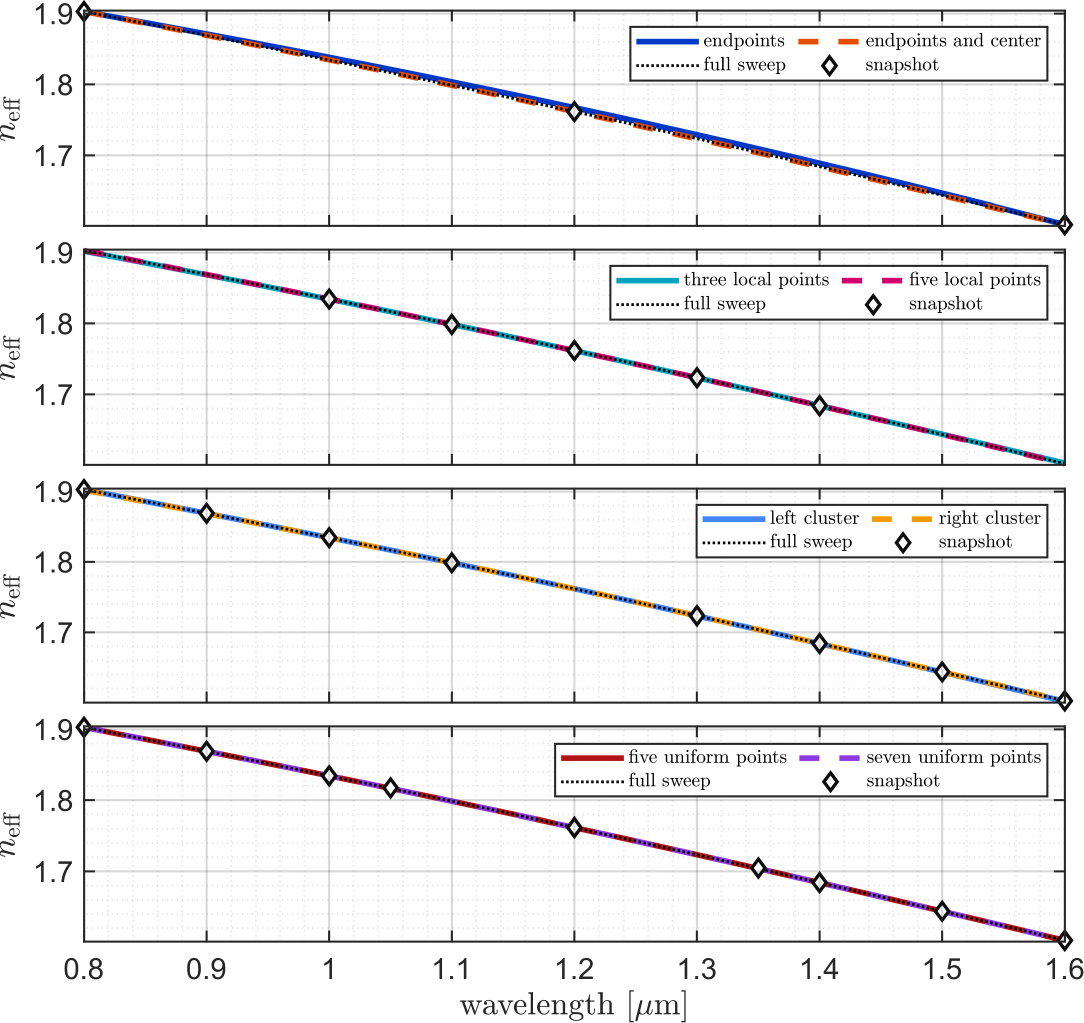}
    \caption{Effective-index reconstruction for different snapshot-placement strategies in the single-mode strip-waveguide wavelength sweep. The black curve denotes the full-order sweep, colored curves denote EC reconstructions, and diamond markers indicate the snapshot wavelengths used to build each reduced basis.}
    \label{Fig2:neff}
\end{figure}

Figure~\ref{Fig2:neff} compares the EC-reconstructed effective-index curves with the full-order wavelength sweep. All snapshot strategies reproduce the global decreasing trend of the fundamental quasi-TE mode. However, the agreement depends strongly on how the snapshots are distributed over the wavelength interval.

The solutions that sample only a restricted region of the interval can accurately represent the modal branch near their training region, but become less reliable away from it. In contrast, uniformly distributed snapshots provide a more stable representation of the modal family over the full wavelength range. This behavior shows that the EC basis does not merely interpolate scalar values of \(n_{\mathrm{eff}}\); rather, it must contain enough field information to span the wavelength-dependent modal manifold.

\subsection{Residual and effective-index error}

To quantify the performance of each basis, we use two complementary diagnostics. The first is the full-operator residual, which measures how well the reconstructed EC field satisfies the full Maxwell eigenproblem at the query wavelength. The second is the absolute effective-index error,

\[
\Delta n_{\mathrm{eff}}(\lambda)
=
\left|n_{\mathrm{eff}}^{\mathrm{EC}}(\lambda)
-
n_{\mathrm{eff}}^{\mathrm{full}}(\lambda)\right|,
\]
which measures agreement with the full-order reference sweep.

\begin{figure}[h]
    \centering
    \includegraphics[width=1.0\linewidth]{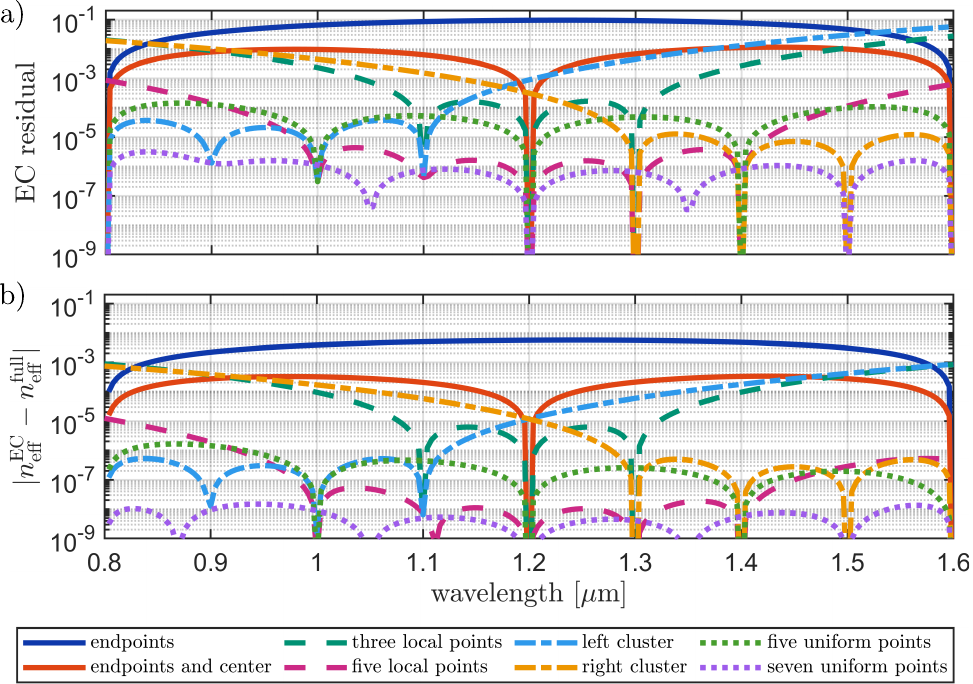}
    \caption{Accuracy diagnostics for the snapshot-placement study. (a) Full-operator residual of the EC solution evaluated at each query wavelength. (b) Absolute effective-index error with respect to the full-order wavelength sweep. Uniformly distributed snapshots provide the most stable representation of the modal branch over the full wavelength interval.}
    \label{Fig3:Residual}
\end{figure}

Figure~\ref{Fig3:Residual} shows both diagnostics. The residual and the effective-index error are related, but they do not measure the same quantity. The effective-index error compares only the eigenvalue prediction with the reference curve, whereas the residual evaluates whether the reconstructed field satisfies the full-order operator at the query point. Therefore, a basis may produce a small effective-index error while still exhibiting a larger residual if the field reconstruction is not fully compatible with the full Maxwell operator.

The best-performing strategies are those that provide low residual and low effective-index error over the entire wavelength interval. In this benchmark, the seven-point uniform basis gives the lowest median residual, \(7.17\times 10^{-7}\), followed by the five-point uniform basis with \(4.60\times 10^{-5}\), and the five-point local basis with \(8.58\times 10^{-4}\). These results show that only five to seven well-distributed snapshots are sufficient to reproduce the full wavelength sweep over that interval with high accuracy.

In the matched tensorial benchmark, the EC online sweep is approximately \(1.5\times\) faster than the full-order sweep. This modest acceleration is explained by the cost distribution: the reduced eigensolve is essentially negligible, while tensor construction and full-operator projection account for approximately \(98\%\) of the online time. A separate four-field implementation produced speedups approaching two orders of magnitude because sparse operator assembly was inexpensive relative to the full eigensolve. Although these results are not directly comparable due to differences in formulation and discretization, they demonstrate that EC acceleration is implementation dependent. Substantial speedups are expected when operator evaluation is inexpensive, cached, affinely decomposed, or hyper-reduced.


This snapshot-placement study shows that the reduced basis must capture the target modal manifold over the complete wavelength interval. Uniformly distributed snapshots provide the most stable residual behavior, whereas localized snapshots lose accuracy away from the training region. This motivates the residual-guided adaptive strategy used in the multimode benchmark: poorly represented wavelengths are identified from the full-operator residual and added to the snapshot basis. 

\section{Multimode EC and Mode-Family Tracking}

After studying EC for a single isolated mode, we now consider a complex multimode problem. In broadband waveguide simulations, several guided modes may coexist over the same wavelength interval. When the full eigenvalue problem is solved independently at each wavelength, the ordering of the computed eigenpairs is not guaranteed to remain fixed. Modes can become closely spaced in effective index, exchange order, or exhibit changes in polarization content and spatial profile. Therefore, a robust reduced model should not only reproduce the effective-index curves, but also preserve the identity of the targeted modal families across the sweep.

This section addresses two questions. First, can a reduced basis containing several modal branches represent a multimode wavelength sweep with high accuracy? Second, can the full-operator residual guide an adaptive enrichment strategy when the initial snapshots are insufficient? To answer these questions, we construct a multimode EC basis from four modes at each selected wavelength and compare the resulting reduced sweeps against a full-order reference solution.

\subsection{Ridge-waveguide benchmark}

We consider a Si\(_3\)N\(_4\) ridge waveguide on a SiO\(_2\) substrate with air cladding. The total core thickness is \(0.90~\mu\mathrm{m}\), including a \(0.15~\mu\mathrm{m}\) slab. The top ridge width is \(1.40~\mu\mathrm{m}\), the slab width is \(3.00~\mu\mathrm{m}\), and the sidewall angle is \(82^\circ\). The computational window is \(4.0 \times 3.2~\mu\mathrm{m}^2\), discretized with \(N_x=379\) and \(N_y=303\) grid points. Dirichlet boundary conditions are imposed at the transverse boundaries, and tensorial subpixel smoothing is used to represent the dielectric interfaces.

\begin{figure}[h]
    \centering
    \includegraphics[width=1\linewidth]{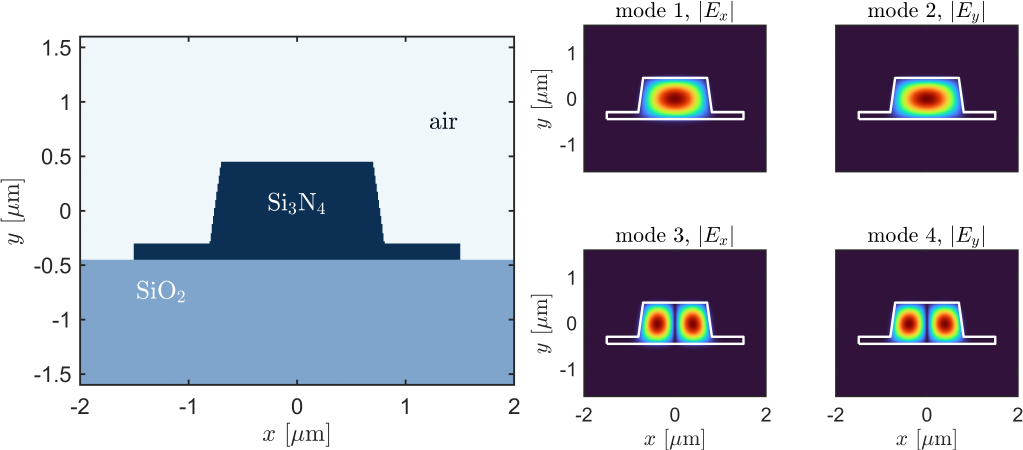}
    \caption{Multimode ridge-waveguide design. Left: Si\(_3\)N\(_4\) ridge waveguide on a SiO\(_2\) substrate with air cladding. Right: field-intensity profiles of the first four guided modes used to construct the multimode EC basis.}
    \label{Fig4:Design2}
\end{figure}

Figure~\ref{Fig4:Design2} shows the ridge cross-section and the four guided modes used in this benchmark. The first four guided modes are tracked over the wavelength interval
\[
\lambda \in [0.40,2.00]~\mu\mathrm{m}.
\]
A complete full-order wavelength sweep is used as the reference solution. This case provides a multimode benchmark for testing whether EC can represent several mode families simultaneously and maintain mode continuity over a broad spectral range.

\subsection{Multimode reduced-basis construction}
In the single-mode benchmark, each snapshot wavelength contributes one eigenvector to the EC basis. In the present multimode case, each snapshot wavelength contributes the first four guided modes. Therefore, for a set of snapshot wavelengths \(\{\lambda_i\}_{i=1}^{n_s}\), the multimode
reduced basis is assembled in blocks as
\begin{equation}
V =
\left[
V(\lambda_1),\,
V(\lambda_2),\,
\ldots,\,
V(\lambda_{n_s})
\right],
\label{eq:multimode_basis}
\end{equation}
where each wavelength block contains the four tracked modes,
\begin{equation}
V(\lambda_i) =
\left[
x_1(\lambda_i),\,
x_2(\lambda_i),\,
x_3(\lambda_i),\,
x_4(\lambda_i)
\right].
\label{eq:multimode_block}
\end{equation}
Here, \(x_m(\lambda_i)\) denotes the full-order eigenvector of mode \(m\) at the snapshot wavelength \(\lambda_i\). For this four-mode benchmark, the number of basis vectors before compression is \(n_b=4n_s\), where \(n_s\) is the number of selected snapshot wavelengths.

\subsection{Adaptive enrichment strategy}

Based on the seed-placement study of the previous section, we initialize the multimode EC model with three different snapshot strategies: five uniformly distributed wavelengths, seven uniformly distributed wavelengths, and four wavelengths locally distributed around the middle of the interval. These initial bases are intentionally different in coverage, allowing us to test whether residual-guided enrichment can compensate for suboptimal initial snapshot placement.

For each initial basis, the EC model is evaluated over the complete wavelength grid. At every query wavelength and for every tracked mode, the full-operator residual is computed. If the maximum residual exceeds the prescribed tolerance, a new full-order solve is performed at the wavelength associated with the largest residual, and the four guided modes at that wavelength are added to the basis. The reduced model is then rebuilt and the process is repeated until the target residual criterion is reached.

In this work, the residual tolerance is set to \(10^{-6}\). Adding a single wavelength snapshot in the multimode case enriches the basis with four eigenvectors, one for each tracked mode. This increases the reduced-basis dimension more rapidly than in the single-mode case, but it also improves the representation of the coupled multimode subspace over the wavelength interval.

\subsection{Tracking accuracy and residual analysis}

\begin{figure}[h]
    \centering
    \includegraphics[width=0.9\linewidth]{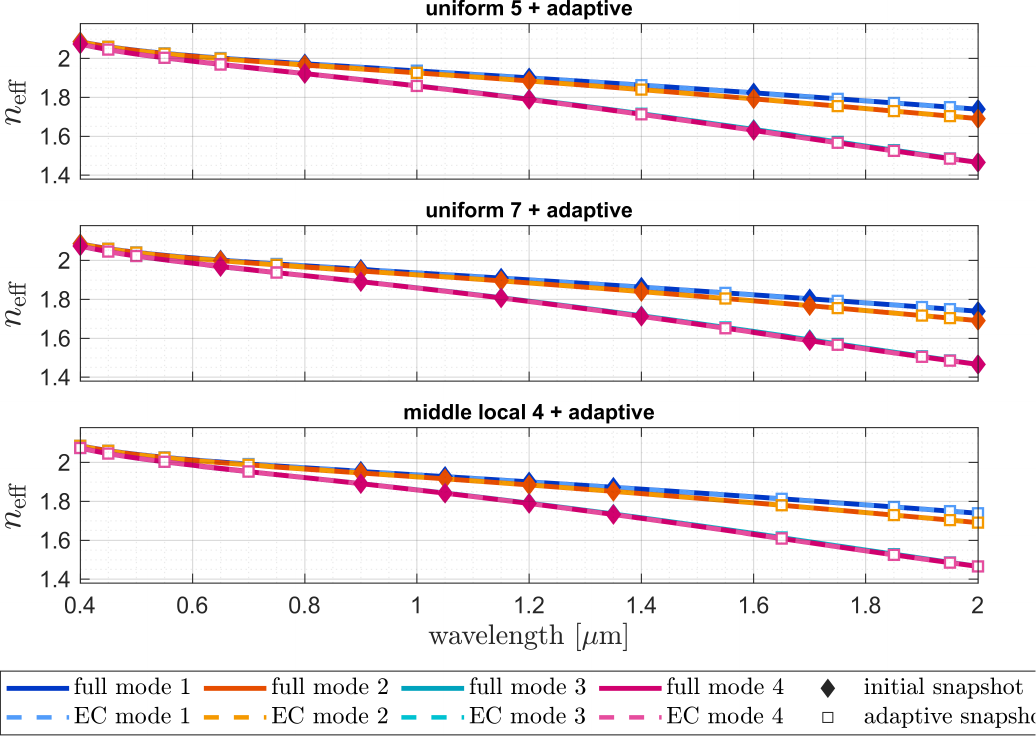}
    \caption{Adaptive multimode EC reconstruction of the first four guided modes of the ridge waveguide. The full-order wavelength sweep is used as reference, while the EC curves are obtained from adaptive reduced bases initialized with different snapshot-placement strategies. Square markers indicate adaptively added snapshot wavelengths.}
    \label{Fig5:neff}
\end{figure}

Figure~\ref{Fig5:neff} compares the EC-reconstructed effective-index curves with the full-order reference sweep for the first four guided modes. All three adaptive strategies recover the full-order modal branches with excellent agreement over the complete wavelength interval. This confirms that the multimode EC basis can represent several mode families simultaneously and can preserve branch continuity even when the modes are computed from a reduced eigenproblem.

Because the reduced space contains snapshots from all targeted modes, the EC reconstruction is guided by modal-field information rather than by effective-index ordering alone. This reduces the fragility of independent eigensolves when branches become close or field profiles evolve across the sweep.

\begin{figure}[h]
    \centering
    \includegraphics[width=1.05\linewidth]{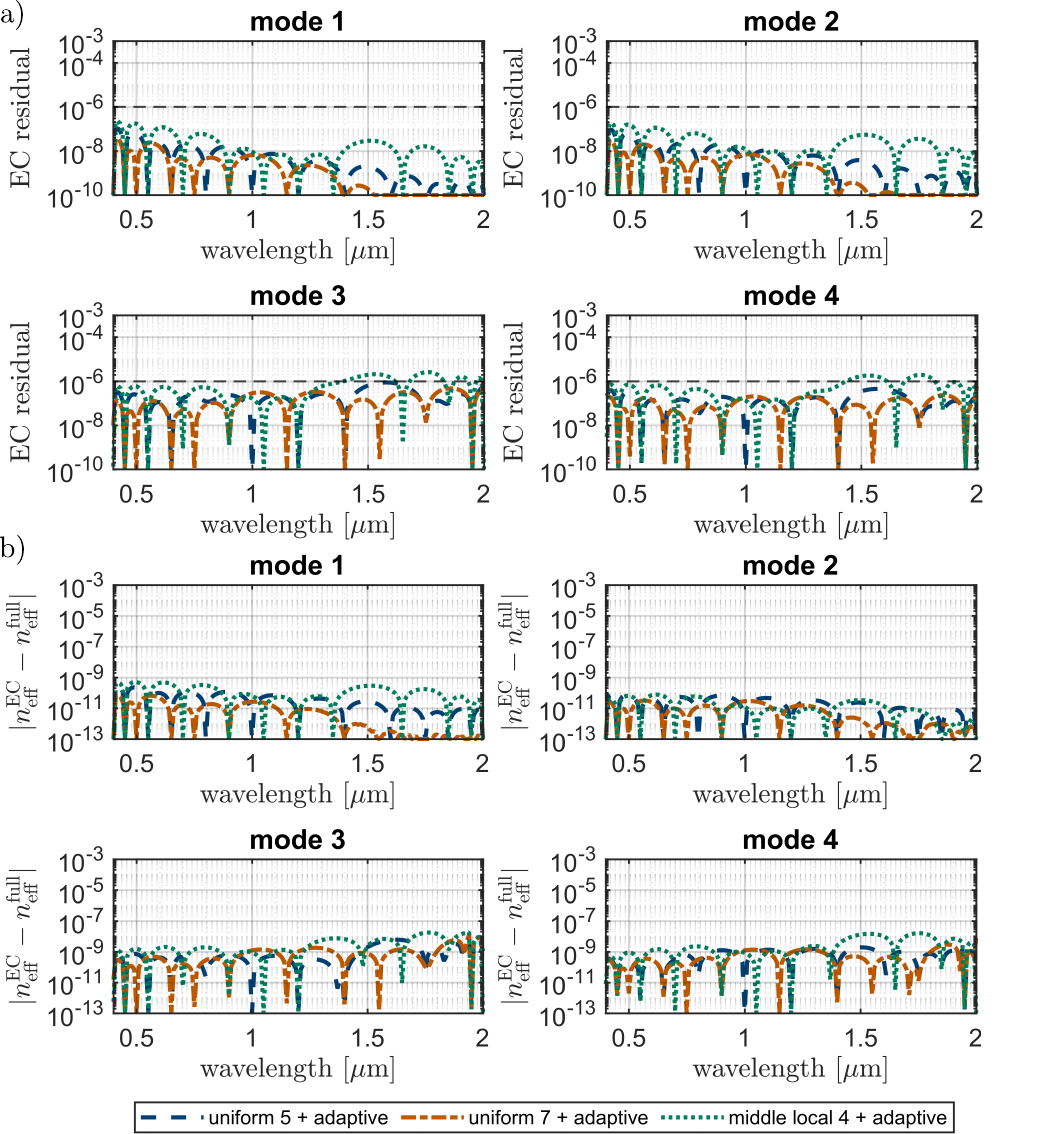}
    \caption{Accuracy diagnostics for the adaptive multimode EC sweep. Panels show the full-operator residual and the absolute effective-index error for the first four guided modes. Residual-guided enrichment reduces the error for all initial snapshot strategies and reaches high-fidelity agreement with the full-order reference sweep.}
    \label{Fig6:Residuals}
\end{figure}

Figure~\ref{Fig6:Residuals} shows the residual and effective-index error for each of the four modes. After residual-guided enrichment, all initial strategies converge to high-accuracy reduced sweeps, with absolute effective-index errors on the order of \(10^{-9}\). The residual therefore acts both as an accuracy indicator and as a sampling criterion for constructing a more informative multimode basis.

\section{Geometry-Dependent Sweeps and Moving-Boundary Effects}

The previous examples considered fixed-geometry wavelength sweeps, where the transverse cross-section remains unchanged and the parameter dependence enters through the wavelength-dependent Maxwell operator. We now consider a more challenging design scenario in which the waveguide geometry itself is varied. This case is important because geometry sweeps are central to photonic design, but they introduce a different numerical difficulty: the dielectric interface moves on the computational grid.

Although the physical waveguide width changes smoothly, the discrete permittivity distribution on a fixed Cartesian grid can change non-smoothly when material boundaries cross grid cells. Therefore, this section evaluates whether EC remains reliable when the parameter dependence enters through the geometry rather than only through \(k_0(\lambda)\) and material dispersion.

\subsection{Top-width sweep benchmark}

We use the same Si\(_3\)N\(_4\) ridge-waveguide platform introduced in the multimode benchmark and vary the top ridge width while keeping the remaining geometrical parameters fixed. The top width is swept over the interval
\[
w_{\mathrm{top}} \in [1.2,1.8]~\mu\mathrm{m}.
\]
At each geometry, the first two guided modes are computed with the full-order solver and used as the reference solution. EC bases are constructed from selected full-order snapshots along the width sweep, and the reduced solutions are compared against the reference sweep using effective index, residual, effective-index error, field overlap, and field-level reconstruction.

This benchmark is designed to distinguish geometry prediction from true low-residual reduced emulation. In particular, we ask whether EC can still provide useful modal predictions when the dielectric interface moves across the computational grid.

\subsection{Effective-index prediction}

\begin{figure}[h]
    \centering
    \includegraphics[width=1.0\linewidth]{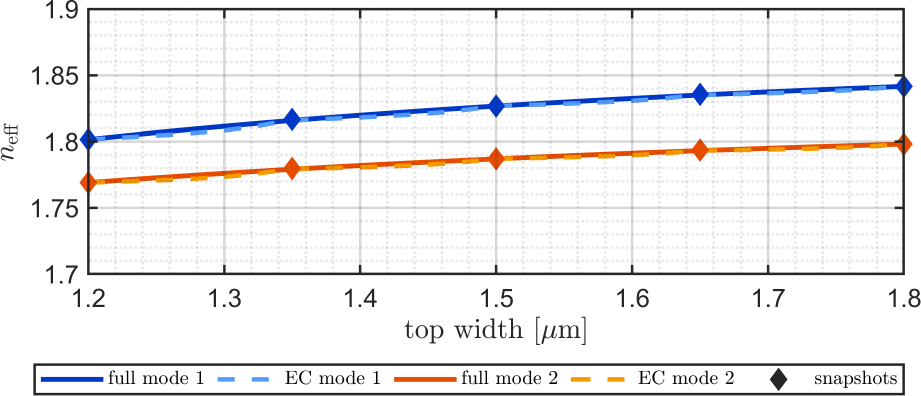}
    \caption{Effective-index prediction for the ridge-waveguide top-width sweep. Solid curves denote the full-order solutions for the first two guided modes, dashed curves denote the EC reconstructions, and diamond markers indicate the geometry snapshots used to construct the reduced basis.}
    \label{Fig7:neff}
\end{figure}

Figure~\ref{Fig7:neff} compares the EC-reconstructed effective indices with the full-order reference sweep for the first two guided modes. The reduced model reproduces the monotonic variation of both modal branches over the full top-width interval. In this sense, EC provides a useful geometry predictor: even though the operator changes with the moving dielectric boundary, the reduced basis still contains enough modal information to approximate the effective-index evolution.

This result is relevant for design workflows because the EC solution can provide accurate initial estimates of \(n_{\mathrm{eff}}\) and can be used to generate improved guesses for subsequent full-order solves. However, agreement in effective index alone is not sufficient to conclude that the reduced solution satisfies the full Maxwell operator at each geometry. For this reason, the residual and field-overlap diagnostics must also be examined.

\subsection{Residual, effective-index error, and field overlap}

\begin{figure}[h]
    \centering
    \includegraphics[width=1.0\linewidth]{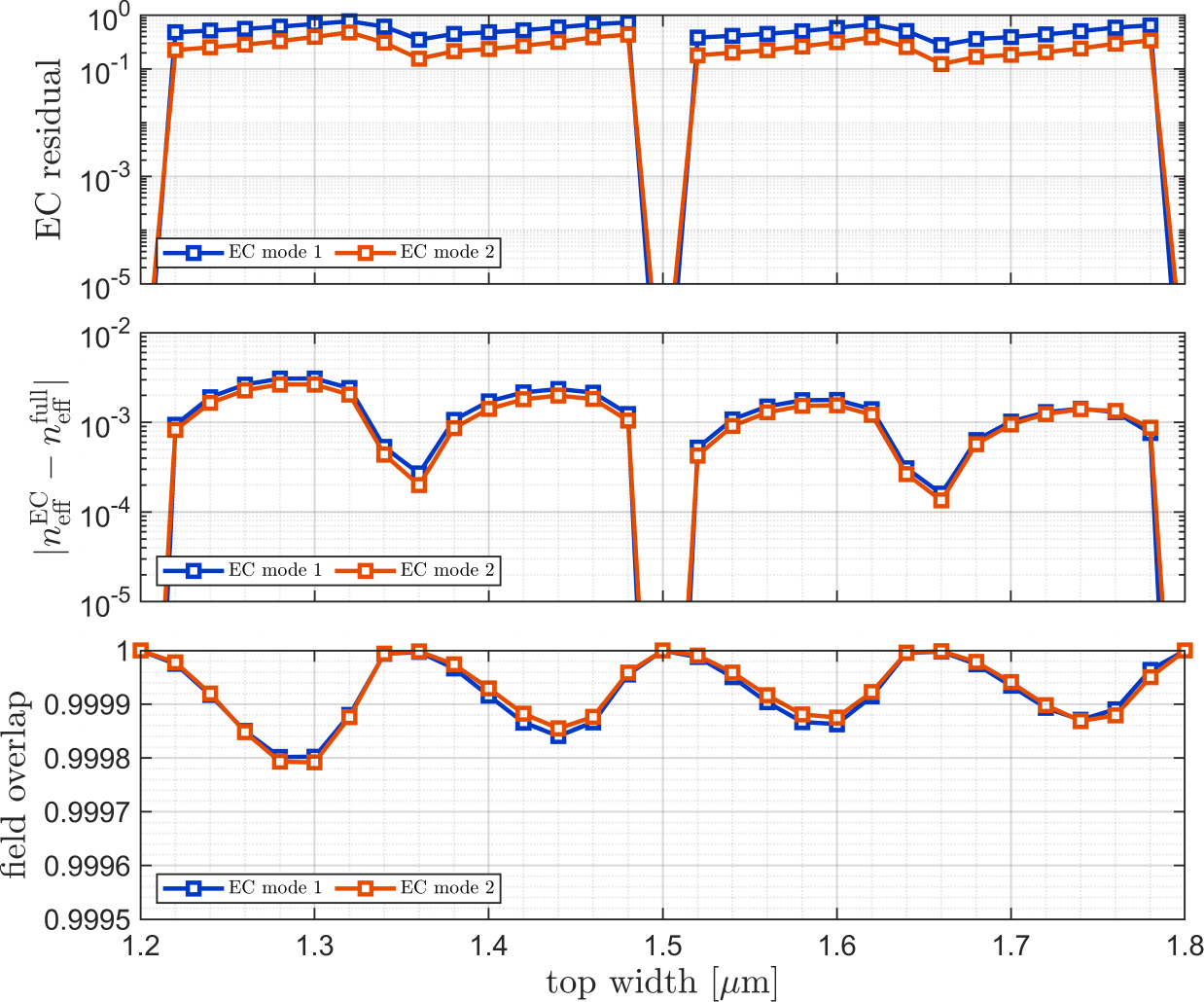}
    \caption{Diagnostics for the top-width geometry sweep. Top: full-operator residual of the EC solution. Middle: absolute effective-index error with respect to the full-order reference solution. Bottom: field overlap between the EC-reconstructed mode and the full-order mode. The reduced model predicts the modal branches and field profiles accurately, but the residual reveals the difficulty of moving-boundary geometry continuation on a fixed Cartesian grid.}
    \label{Fig8:residuals}
\end{figure}

Figure~\ref{Fig8:residuals} shows three complementary diagnostics for the same width sweep: the full-operator residual, the absolute effective-index error, and the field overlap between the EC-reconstructed field and the full-order reference field. The effective-index error remains small (orders of $\approx$ \(10^{-3}\)) over the sweep, and the field overlap stays close to unity for both modes. This indicates that the reduced model captures the dominant modal shape and provides an accurate prediction of the modal branch.

In contrast, the full-operator residual is significantly larger than in the fixed-geometry wavelength-sweep cases. This apparent discrepancy is important. The effective-index error measures agreement with the reference eigenvalue, and the overlap measures similarity between field profiles. The residual, however, measures whether the reconstructed field satisfies the target discrete Maxwell operator at the new geometry. Therefore, a reduced field can have high overlap with the full-order mode and still produce a large residual if the discrete operator changes non-smoothly with the geometry parameter.

This behavior suggests that, for moving-boundary sweeps on fixed Cartesian grids, EC remains valuable as a predictor but does not yet behave as a fully low-residual reduced solver. The residual exposes the mismatch between a smoothly varying physical geometry and a discretized operator whose material distribution can change abruptly as the boundary moves across grid cells.

\subsection{Field reconstruction and moving-boundary error}

\begin{figure}[h]
    \centering
    \includegraphics[width=1.0\linewidth]{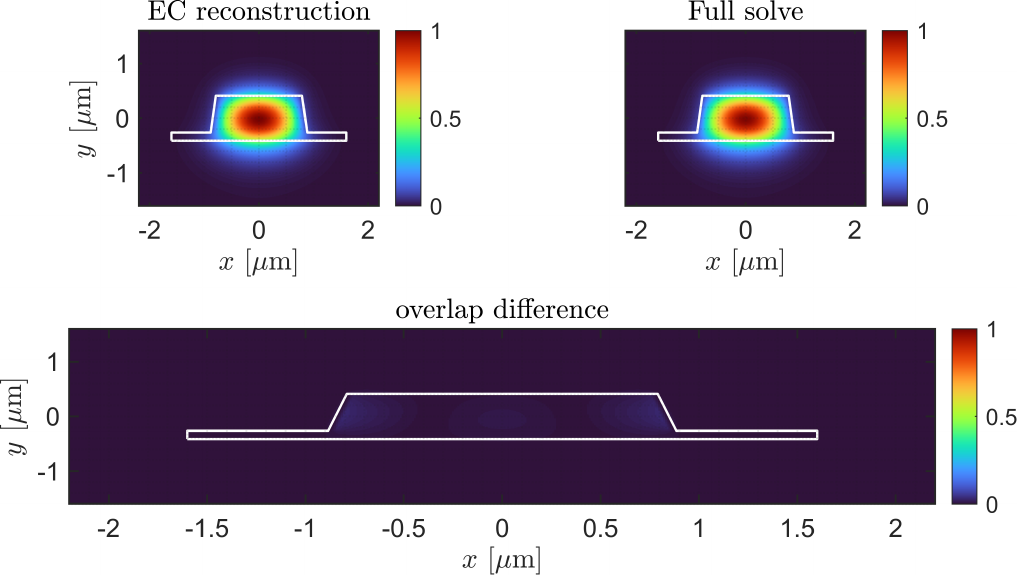}
    \caption{Field-level comparison for a representative point in the top-width sweep. Top left: normalized dominant field component reconstructed by EC. Top right: normalized dominant field component obtained from the full-order solve. Bottom: normalized field-difference map. The EC and full-order fields are visually similar, while the remaining discrepancy is localized mainly near dielectric interfaces, supporting the interpretation that the large residual is dominated by moving-boundary discretization effects.}
    \label{Fig9:fields}
\end{figure}

To further interpret the residual behavior, Figure~\ref{Fig9:fields} compares the EC-reconstructed field with the corresponding full-order field at a representative width. The reconstructed dominant field component is visually almost identical to the full-order solution, consistent with the high field overlap observed in Figure~\ref{Fig8:residuals}. The difference map shows that the remaining discrepancy is mainly localized near the dielectric interfaces, especially around the ridge sidewalls and slab discontinuities.

This localization supports the interpretation that the dominant error is associated with the moving material boundary rather than with a failure to represent the global modal shape. In a fixed-grid geometry sweep, small changes in the physical boundary can modify the local dielectric assignment near the interface. Even when tensorial subpixel smoothing is used~\cite{Farjadpour2006,Kottke2008}, the discrete operator may not vary as smoothly as the underlying physical geometry. Consequently, the EC basis can predict the modal field and effective index accurately while still producing a relatively large operator residual.

\section{Discussion}

The three examples define the practical scope of EC-based photonic mode emulation. For fixed-geometry wavelength sweeps, the discrete Maxwell operator varies smoothly with wavelength, and a small number of well-distributed snapshots can span the target modal family. In this regime, EC behaves as an operator-consistent reduced modal solver: it reconstructs fields, predicts propagation constants, and provides a residual-based reliability indicator.

For multimode sweeps, the main benefit is not only reduced dimension but modal organization. By embedding several modal families in a shared reduced basis, EC reduces the dependence on effective-index sorting and supports robust mode-family tracking. The residual also provides a natural adaptive sampling criterion. In practice, batch enrichment could further reduce repeated adaptive sweeps by adding several high-residual wavelengths per iteration.

Geometry-dependent sweeps reveal a different limitation. The modal fields and effective indices can remain smooth, while the fixed-grid operator changes non-smoothly as dielectric interfaces move across cells. In this regime, EC remains useful as a predictor and as a source of initial guesses for full-order solves, but low-residual emulation requires smoother parameterized operators, local EC bases, reference-domain mappings, or hyper-reduced formulations.

\section{Conclusion}

We presented an adaptive eigenvector-continuation framework for full-vector photonic waveguide eigenmode sweeps. The method builds a reduced basis from selected full-order modal snapshots, solves projected Maxwell eigenproblems at new query points, reconstructs modal fields, and evaluates a full-operator residual as a reliability indicator.

The numerical results show that EC is effective for fixed-geometry wavelength sweeps, where well-distributed snapshots reproduce the target modal branch with low residual and low effective-index error. In a multimode ridge waveguide, a shared reduced basis enables robust broadband mode-family tracking and residual-guided enrichment. In geometry-dependent width sweeps, EC provides accurate effective-index predictions and high-overlap field reconstructions, while the residual exposes moving-boundary errors associated with fixed-grid discretizations.

These results suggest that adaptive EC is best understood as an operator-consistent modal emulator and diagnostic tool for photonic waveguide sweeps. Future work will focus on faster online operator projection and geometry-aware formulations for low-residual reduced emulation of parameterized photonic structures.


\bibliographystyle{IEEEtran}
\bibliography{refs}

\end{document}